\documentclass[epj]{webofc}
\usepackage[varg]{txfonts}   
\usepackage{caption}
\usepackage{wrapfig}
\usepackage{subcaption}
\usepackage{algorithm}
\usepackage{algpseudocode}
\usepackage{color}
\usepackage{soul}
\def\bareta{\eta\kern-0.8ex\lower0.2ex\hbox{\vrule height0.4pt width0.8ex}\kern0.1ex}
\woctitle{Mathematical Modeling and Computational Physics 2015}

\usepackage[english]{babel}
\begin{document}
\title{Algorithm for Solving the Optimization Problem\\ for the Temperature Distribution on a Plate}
%
%

\author{A.~Ayriyan\inst{1}\fnsep\thanks{\email{ayriyan@jinr.ru}} \and
        E.E.~Donets\inst{1} \and H.~Grigorian\inst{1,2} \and N.~Kolkovska\inst{3} \and A.~Lebedev\inst{1,4}
}

\institute{
		Joint Institute for Nuclear Research, Joliot-Curie 6, 141980 Dubna, Moscow Region, Russia
\and
		Yerevan State University, Alek Manyukyan 1, 0025 Yerevan, Republic of Armenia
\and 
		Institute of Mathematics and Informatics of BAS, Acad. Georgi Bonchev 8, 1113 Sofia, Bulgaria
\and
		GSI Helmholtzzentrum f\"{u}r Schwerionenforschung, Planckstra{\ss}e 1, 64291 Darmstadt, Germany
          }

\abstract{%
The work describes the maximization problem regarding heating of an area on the surface of a thin plate within a given temperature range. The~solution of the problem is applied to ion injectors. The given temperature range corresponds to a required pressure of a saturated gas comprising evaporated atoms of the plate material. In order to find the solution, a one-parameter optimization problem was formulated and implemented leading to optimization of the plate's specific geometry. It was shown that a heated area can be increased up to $23.5\%$ in comparison with the regular rectangle form of a given plate configuration.
}
\maketitle
%

\section{Introduction}
\label{intro}

The work describes the maximization problem regarding heating of an area on the surface of a thin plate within a given temperature range. The plate serves for injecting the working species (atoms of the plate material evaporated from its surface) into the working space of the ion source~\cite{donets_2010}.
The plate is heated by the flux of electric current passing through it. The injection starts when the temperature reaches the required value depending on the material of the plate. The temperature range and the working area of the surface respectively are defined by the required pressure of a saturated vapor above the surface of the plate. 

\begin{figure}[h!]
    \centering
    \begin{subfigure}[b]{0.49\textwidth}
        \centering
        \includegraphics[width=1.0\textwidth]{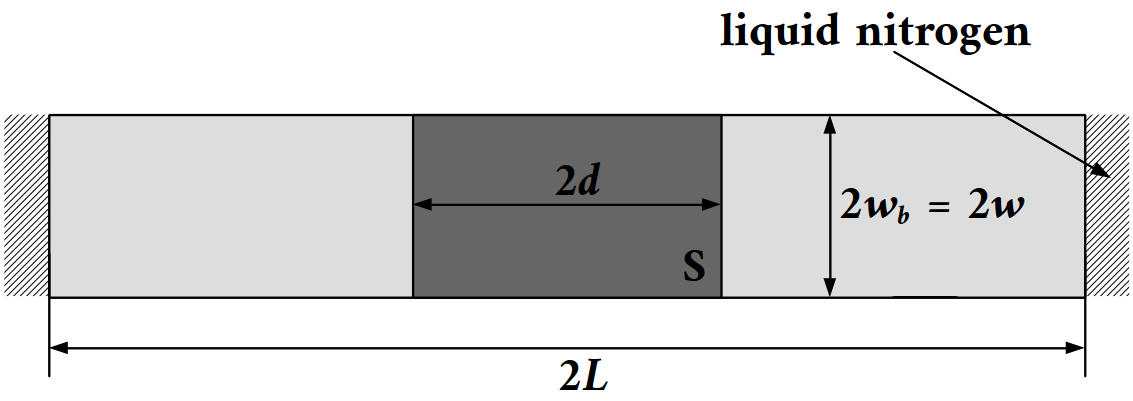}
        \caption{Sketch of the existing plate geometry.}
        \label{fig:existing_plate}
    \end{subfigure}
    \hfill
    \begin{subfigure}[b]{0.49\textwidth}
        \centering
        \includegraphics[width=1.0\textwidth]{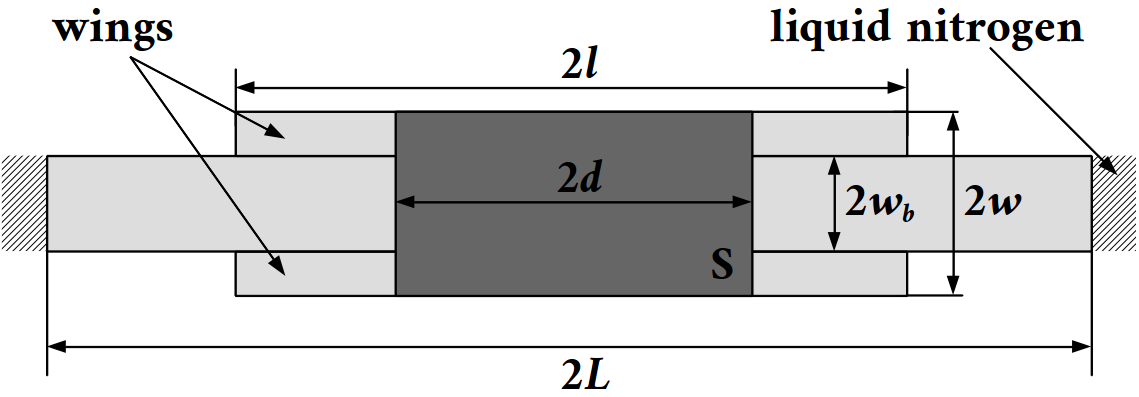}
        \caption{Sketch of the suggested geometry.}
        \label{fig:suggested_plate}
    \end{subfigure}
    \vspace{-5mm}
    \caption{The geometry of the plate. Here $2L$ is a length of the plate, $2w$ is a width of the plate, $2l$ is a length of the wings, $2w_b$ is a width without wings, $S$ is a working area of the plate surface heated to the required temperatures, $2d$ is a length of the working area. Left and right sides are connected to the temperature terminal.}
    \label{fig:object}
    \vspace{-5mm}
\end{figure}

In this work a model of the plate and a one-parameter variation of its geometry are discussed (see Fig.~\ref{fig:object}). In the existing technical device the plate has a rectangular form (Fig.~\ref{fig:existing_plate}). In order to maximize the working area on its surface we suggested to change the geometry in the following way. A new shape has been derived from the regular one by removal of rectangular parts from the corners. Thus rectangular wing-like structures (further referenced as {\it wings}) appear on the both sides of the plate (Fig.~\ref{fig:suggested_plate}) with their length being specified by us as a free parameter. Our simulations have shown that the working area of the plate has approximately a rectangular shape for the used set of parameters. The shape of the plate has been optimized by varying the length of these wings in order to reach a maximum of the working area. The procedure has been applied under the condition that the highest temperature has to be significantly less than the melting temperature of the plate material.

\section{Main Equation and Conditions}
\label{main}

The stationary temperature distribution in the plate could be modelled by the following equation \cite{samarski_1995}:
\begin{equation}
\label{eq:main_equation}
\frac{\partial}{\partial x} \left( \lambda(T) \frac{\partial T}{\partial x}\right ) + \frac{\partial}{\partial z} \left(\lambda(T) \frac{\partial T}{\partial z}\right) + \dfrac{I^2\chi(T)}{S^2_{\mathrm{C}}} = 0,
\end{equation}
where the thermal conductivity $\lambda(T)$ and the resistivity $\chi(T)$ are non-linearly dependent on a sought-after function (temperature), $S_{\mathrm{C}}$ is a cross-sectional area, and $I$ is an electric current.

Due to the symmetry at the middle of the plate along the $x$- and $z$- axes solving the problem on a quarter of the full domain is sufficient (see Fig.~\ref{fig:domain}):


\begin{wrapfigure}{r}{0.4\textwidth}
	\vspace{-10mm}
	\begin{center}
		\includegraphics[height=2.0cm]{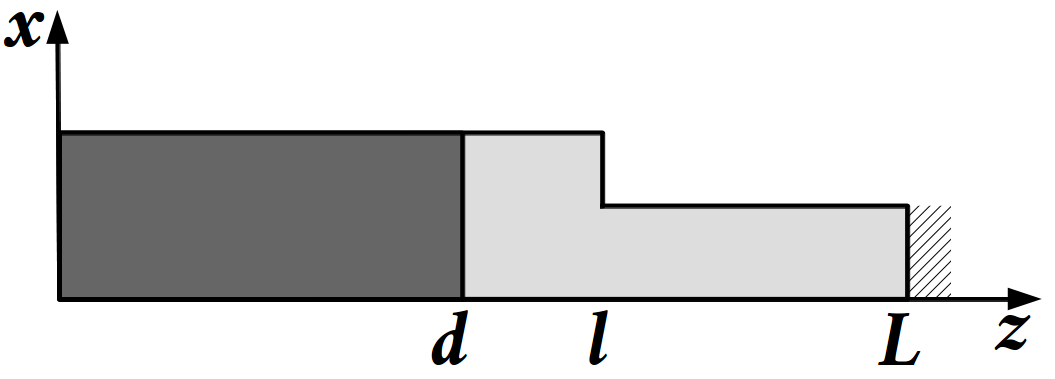}
	\end{center}
	\vspace{-5mm}
	\caption{$\Omega$ -- domain of simulations.}
	\vspace{-8mm}
	\label{fig:domain}
\end{wrapfigure}

The boundary conditions can be taken as following:
\begin{equation}
\label{eq:boundary_cond}
\begin{cases}
\dfrac{\partial T}{\partial \textbf{n}} = 0  & \mbox{if } (x,z) \in \partial\Omega \textbackslash \{(x,z) \vert z=L\}, \\
T = T_{0} & \mbox{if } (x,z) \in \{(x,z) \vert z=L\}.
\end{cases}
\end{equation}
The right side of the plate is connected to the temperature terminal with $T_{0} = 78~\mathrm{K}$. At the boundary of the domain the temperature flow equals to zero: along the $x$- and $z$- axes because of the symmetry and at~the other points because the plate is placed inside the vacuum chamber.

The problem (\ref{eq:main_equation})--(\ref{eq:boundary_cond}) can be solved by various methods, we have chosen the method considering a solution of the elliptic equation (\ref{eq:main_equation}) as a stationary solution of the parabolic one (\ref{eq:equation}):
\begin{equation}
\label{eq:equation}
\rho(T) {c_V}(T)\frac{\partial T}{\partial t}=\frac{\partial}{\partial x} \left( \lambda(T) \frac{\partial T}{\partial x}\right ) + \frac{\partial}{\partial z} \left(\lambda(T) \frac{\partial T}{\partial z}\right) + I^2 X(T),
\end{equation}
here the density is $\rho(T)$, the heat capacity is $c_V(T)$, and $X(T) = \chi(T)/S^2_{\mathrm{C}}$. The solution of the elliptic equation (\ref{eq:main_equation}) is the stationary solution of the heat equation (\ref{eq:equation}). In order to solve the equation (\ref{eq:equation}) we have chosen the initial condition to be:
\begin{equation}
\label{eq:initial_cond}
T = T_{0} \quad \forall \: (x,z) \in \Omega.
\end{equation}

\section{Formulation of Optimization Problem}
\label{intro}
The optimization problem has been formulated to~maximize the working area. Temperature values range from $T_{\textrm{low}}=678~\mathrm{K}$ to $T_{\textrm{high}}=778~\mathrm{K}$. $T_{\textrm{high}}$ has been chosen to be significantly less than the melting point of the plate material. The maximum temperature constraint is applied to the hottest point on the surface of the plate which is located at its center due to plate's homogeneity and symmetry.

\begin{wrapfigure}{r}{0.4\textwidth}
	\vspace{-3mm}
	\begin{center}
		\includegraphics[width=0.4\textwidth]{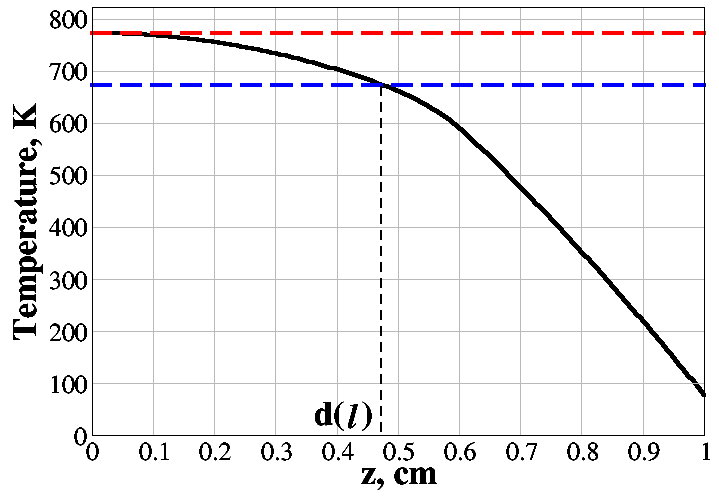}
	\end{center}
	\vspace{-5mm}
	\caption{A typical temperature profile in dependence on $z$ for different length of the wings at $x=0$. The dashed lines correspond to $T_{\textrm{low}}=678~\mathrm{K}$ and $T_{\textrm{high}}=778~\mathrm{K}$.}
	\label{fig:d_def}
\end{wrapfigure}

The size of the working area is proportional to $d$. Thus the maximum of the working area is reached for the maximum of $d$. Note, that the width of the plate $2w$ is a constant and has a maximum possible value based on the technical requirements. The relation $w_b/w$ is fixed. 
Being dependent on the semilength of the wings $l$ (see Fig.~\ref{fig:d_def}) for the given temperature $T_{\mathrm{low}}$, $d$ is the solution of the following equation for $z$:
\begin{equation}
\label{eq:d_eq}
T(x=0,z;l)-T_{\mathrm{low}}=0,
\end{equation}
when we choose the source coefficient $I$ (electrical current) satisfying the maximum temperature constraint.

Therefore, the maximum of the working surface area $S$ corresponds to the maximum of $d(l)$:
\begin{equation}
\label{eq:d_param}
d\left(l^*\right) = \mathop{\max}_{0 \leq l \leq L} |d\left(l\right)|.
\end{equation}

An implementation of the optimization problem requires solving of the heat conduction problem (\ref{eq:boundary_cond})--(\ref{eq:initial_cond}) numerous times.


\section{Numerical Algorithms}
\label{num_alg}
\subsection{Solving of Direct Problem}

The finite difference method has been chosen to solve the heat conduction problem (\ref{eq:boundary_cond})--(\ref{eq:initial_cond}):
\begin{equation}
\label{eq:diff_equation}
\rho_{i,j}\,{c_V}_{i,j}\dfrac{\widehat{T}_{i,j}-{T}_{i,j}}{\tau} = {\textrm{\Large{$\Lambda_x$}}} \left(\widehat{T}_{i,j}\right)+{\textrm{\Large{$\Lambda_z$}}} \left(T_{i,j}\right)+I^2{X}_{i,j},
\end{equation}
\begin{equation}
\label{eq:Lambda_ij}
\begin{cases}
{\textrm{\Large{$\Lambda_x$}}}\left[\right.\widehat{T}_{i,j}\left.\right]=
\frac{1}{\hbar_{i}}\left[\lambda_{i+\frac{1}{2},j} 
\frac{\widehat{T}_{i+1,j}-\widehat{T}_{i,j}}{h_{i+1}} - \lambda_{i-\frac{1}{2},j}
\frac{\widehat{T}_{i,j}-\widehat{T}_{i-1,j}}{h_{i}}\right],
\\[2mm]
{\textrm{\Large{$\Lambda_z$}}}\left[\right.T_{i,j}\left.\right]=
\frac{1}{\bareta_j}\left[\lambda_{i,j+\frac{1}{2}} 
\frac{T_{i,j+1}-T_{i,j}}{\eta_{j+1}}-\lambda_{i,j-\frac{1}{2}}
\frac{T_{i,j}-T_{i,j-1}}{\eta_{j}}\right],
\end{cases}
\end{equation}
where $i = 1 \ldots N_j-1$, $j = 1 \ldots M_i-1$, $h_{i} = x_{i}-x_{i-1}$, $\eta_{j} = z_{j}-z_{j-1}$, $\hbar_{i}=\left(h_{i+1}+h_{i}\right)/2$,

\hspace{-6mm} $\bareta_j=\left(\eta_{j+1}+\eta_{j}\right)/2$, $\displaystyle T_{i,j}=T(x_i,z_j,t_k)$, $\displaystyle \widehat{T}_{i,j}=T(x_i,z_j,t_{k+1})$, $\displaystyle {c_V}_{i,j}={c_V}(T_{i,j})$, $\displaystyle X_{i,j}=X(T_{i,j})$,


\hspace{-6mm} $\lambda_{i\pm \frac{1}{2},j}=\lambda \left( \left( T_{i,j}+T_{i\pm1,j} \right) / 2 \right)$, $\lambda_{i,j\pm \frac{1}{2}}=\lambda \left( \left( T_{i,j}+T_{i,j\pm1} \right) / 2 \right)$.
\vspace{3mm}

The difference scheme (\ref{eq:diff_equation})--(\ref{eq:Lambda_ij}) has been solved by Thomas algorithm \cite{thomas_1949, press_2007}:
\begin{equation}
\label{eq:Thomas_alg}
\alpha_{i} = \dfrac{-C_{i}}{B_{i}+A_{i}\alpha_{i-1}},
\qquad
\beta_{i} = \dfrac{F_{i}-A_{i}\beta_{i-1}}{B_{i}+A_{i}\alpha_{i-1}},
\qquad
\widehat{T}_{i,j} = \alpha_i\widehat{T}_{i+1,j} + \beta_i.
\end{equation}

The coefficients $A_i$, $B_i$, $C_i$, and $F_i$ are defined from the difference equation~(\ref{eq:diff_equation}):
\begin{equation}
\label{ABC}
\left\{
\begin{array}{l}
\displaystyle A_{i} = -\frac{\lambda_{i-\frac{1}{2},j}}{\hbar_{i}\,h_{i}},\\[3mm]
\displaystyle B_{i} = \frac{1}{\hbar_{i}}\left[ \frac{\lambda_{i-\frac{1}{2},j}}{h_{i}} + \frac{\lambda_{i+\frac{1}{2},j}}{h_{i+1}} \right] + \frac{\rho_{i,j}\,{c_V}_{i,j}}{\tau},\\[3mm]
\displaystyle C_{i} = -\frac{\lambda_{i+\frac{1}{2},j}}{\hbar_{i}\,h_{i+1}},\\[3mm]
\displaystyle F_{i} = \frac{\rho_{i,j}\,{c_V}_{i,j}}{\tau}T_{i,j} + \Lambda_z\left[\right.T_{i,j}\left.\right] + I^2X_{i,j}.\\[1mm]
\end{array}
\right.
\end{equation}

The boundary conditions (\ref{eq:boundary_cond}) in Thomas algorithm could be approximated as following:
\begin{equation}
\label{eq:diff_bc_begin}
\alpha_0 = \dfrac{2\lambda_{0,j}\tau}{h_1^2\rho_{0,j}{c_V}_{0,j}+2\lambda_{0,j}\tau},
\qquad
\beta_0  = \dfrac{\rho_{0,j}{c_V}_{0,j}T_{0,j}+\tau\left(\Lambda_z\left[T_{0,j}\right]+I^2X_{0,j}\right)}{h_1^2\rho_{0,j}{c_V}_{0,j}+2\lambda_{0,j}\tau} h_{1}^2,
\end{equation}
\begin{equation}
\label{eq:dif_bc_end}
\widehat{T}_{N_j,j} = \frac{\left[4-\alpha_{N_j-2}\right]\beta_{N_j-1}-\beta_{N_j-2}}{3-\alpha_{N_j-1}\left[4-\alpha_{N_j-2}\right]}.
\end{equation}
Corresponding to (\ref{eq:initial_cond}) the initial condition is $T_{i,j}=T_{0}\:\forall\:i,j$.

The difference scheme (\ref{eq:diff_equation})--(\ref{eq:dif_bc_end}) is the second order approximation of the heat conduction problem (\ref{eq:boundary_cond})--(\ref{eq:initial_cond}) by spatial steps while being the first order by time-step.
The scheme is unconditionally stable relating to spatial step $h_{i}$ and conditionally stable relating to $\eta_{j}$ \cite{yanenko_1967}:
\begin{equation}
\displaystyle \tau \le \frac{\min\left| \eta^2_j \right|}{2}\cdot \min\left| \frac{\rho(T) {c_V}(T)}{\lambda(T)} \right|.
\label{stability_condition}
\end{equation}

\subsection{Algorithm for Solving Optimization Problem}
\label{optimal_control_problem}

The algorithm comprises the following steps: 1) the length of wings $l\in\left[0,L\right]$ is sampled with a number $N_{\mathrm{wings}}$; 2) for each $l_i$ the source problem is solved and $d(l_i)$ is calculated; 3) the value of $l^*$ is found according to the maximum value of $d$.

The value of the electric current $I$ varies in the range of $\left[0,I_{\mathrm{max}}=500~\mathrm{mA}\right]$. 
If $I$ equals to $I_{\mathrm{max}}$, the temperature in the central point of the plate exceeds the maximum temperature constraint for all $l\in\left[0,L\right]$. In order to find the only solution existing within the defined range of $I$, the bisection method is used.

The algorithm for solving the optimization problem is described by the pseudo code 
following below.
\begin{algorithm}
\renewcommand{\algorithmicrequire}{\textbf{\quad\,\, Input:}}
\renewcommand{\algorithmicensure}{\textbf{\quad\,\, Output:}}
\caption{The algorithm for Solving Optimization Problem (\ref{eq:d_param})}
\begin{algorithmic}[1]
\Procedure {SolveOCP}{}
\Comment{Solve (\ref{eq:d_param})}
  \Require $\mathbb{L}$ -- the finite set of $l_{i}$ values, where $i=0\dots N_{\mathrm{wings}}-1$
  \Ensure A value of $l_{i}$ corresponding to $\sup\left(\mathbb{D}\right)$ (maximal value of parameter $d$)
  \ForAll {$l_i \in \mathbf{L}$}
    \State $d_i =$ SolveSCP$\left(l_i\right)$
    \Comment{See lines 8--18}
  \EndFor
  \State $ind = $ FindMax$\left(\mathbb{D}\right)$\\
  \Return $l_{ind}$ and $d_{ind}$
\EndProcedure
\Statex
\Procedure {SolveSCP}{$l_i$, ${I}_{\mathrm{min}}$, ${I}_{\mathrm{max}}$}
  \Require $l_i$, ${I}_{\mathrm{min}}=0$ and ${I}_{\mathrm{max}}=500$
  \Ensure A value of $d_i$ for corresponding value of $l_i$
  \While {$\vert T(0,0,l)-T_{\textrm{high}}\vert \geq \varepsilon$}
    \State $I = \left({I}_{\mathrm{min}} + {I}_{\mathrm{max}}\right) / 2$
    \State $T\left(x,z,l_i\right) = $ SolveDP($I$,$l_i$)
    \Comment{Solve direct problem for heat equation (\ref{eq:boundary_cond})--(\ref{eq:initial_cond})}
    \If{$T(0,0,l_i) > T_{\textrm{high}}$}
      \quad ${I}_{\mathrm{max}}=I$
    \Else
      \quad ${I}_{\mathrm{min}}=I$
    \EndIf
  \EndWhile
    \State $d_i =$ SolveEq\ref{eq:d_eq}$\left(T\left(0,z,l_i\right)\right)$
    \Comment{Solve equation (\ref{eq:d_eq})}\\
    \Return $d_i$
\EndProcedure
\end{algorithmic}
\label{alg:SolveOCP}
\end{algorithm}
\vspace{-3mm}

\subsection{Parallel Algorithm}
\label{parallel_alg}

The parallel algorithm was implemented with the usage of Message Passing Interface (MPI) \cite{open_mpi}. The main loop of the algorithm~\ref{alg:SolveOCP} (line~2) was parallelized. For each $l_i$ the source problem is solved by a separate parallel process. The parallel algorithm may be executed up to $N_{\mathrm{wings}}$ times faster in comparison with the non-parallel one.


\section{Results}
\label{results}

The results are presented for the plate made of Thulium. Generally, the algorithm has been developed for temperature values depending on thermal coefficients, but for the present simulations the thermal coefficients \cite{handbook, thulium} are assumed to be constant for the considered temperature range: the conductivity is $\lambda = 0.169~\mathrm{J/(cm~s~K)}$, the specific heat is $c_V = 0.16~\mathrm{J/(g~K)}$, the density is $\rho = 9.33~\mathrm{g/{cm^3}}$, and the ratio resistivity to square of the cross-section area is considered to be the same throughout the whole plate $\chi/S^2_{\mathrm{C}} = 10^3 \times 10^{3}~\mathrm{Ohm / cm^3}$. The fixed size of the plate is the following $L=1~\mathrm{cm}$, $w_{b}=0.05~\mathrm{cm}$ and $w=0.125~\mathrm{cm}$ (see Fig.~\ref{fig:suggested_plate}). The~sampling number $N_{\mathrm{wings}}=10$.

\begin{figure}[h!]
    \centering
    \begin{subfigure}[b]{0.31\textwidth}
        \centering
        \includegraphics[width=1.1\textwidth]{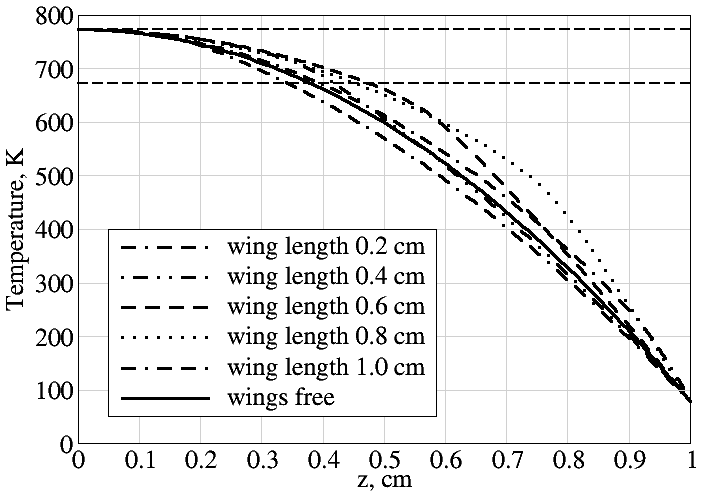}
        \caption{Temperature profiles for different lengths of wings}
        \label{fig:temperature_profiles}
    \end{subfigure}
    \hfill
    \begin{subfigure}[b]{0.31\textwidth}
        \centering
        \includegraphics[width=1.1\textwidth]{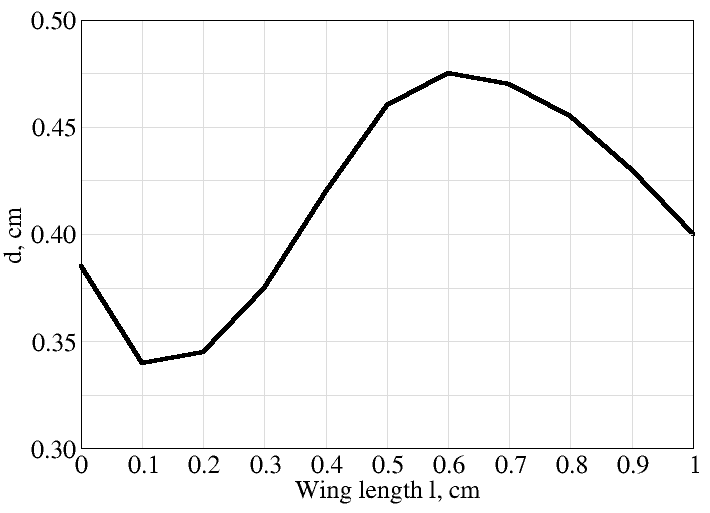}
        \caption{Parameter $d$ for different length of wings}
        \label{fig:d_parameter}
    \end{subfigure}
    \hfill
    \begin{subfigure}[b]{0.31\textwidth}
        \centering
        \includegraphics[width=1.2\textwidth]{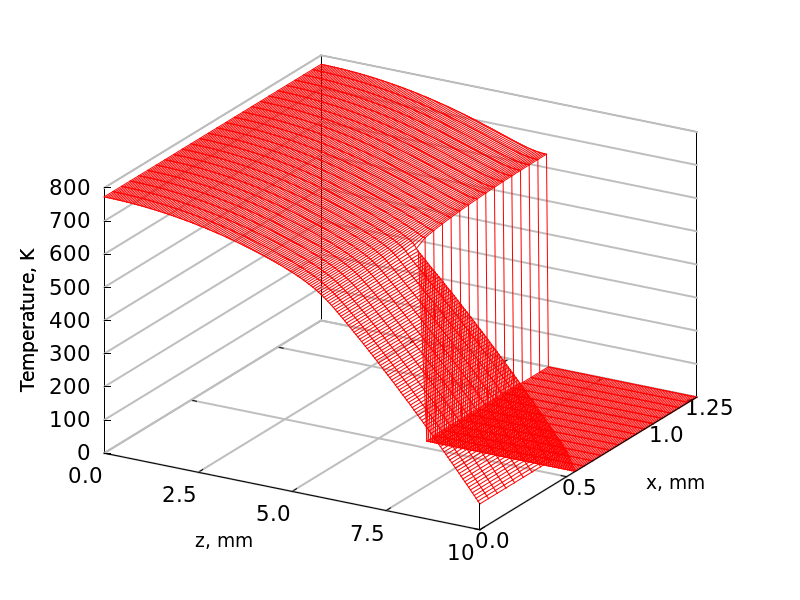}
        \caption{Temperature field in quota of~the~plate}
        \label{fig:temperature_field}
    \end{subfigure}
    \vspace{-5mm}
    \caption{Results of solving the source problem as a function of $l$}
    \label{fig:main_results}
\end{figure}

One can see from the Fig.~\ref{fig:temperature_profiles} that the temperature profile is sensitive to the length of the wings. The function $d\left(l\right)$ and the temperature field for $l^*$ are shown in Fig.~\ref{fig:d_parameter} and Fig.~\ref{fig:temperature_field} respectively. The maximum $d$ corresponds to $l^*=0.6~\mathrm{cm}$.

\begin{figure}[h!]
    \vspace{-4mm}
    \centering
    \begin{subfigure}[b]{0.32\textwidth}
        \centering
        \includegraphics[width=\textwidth]{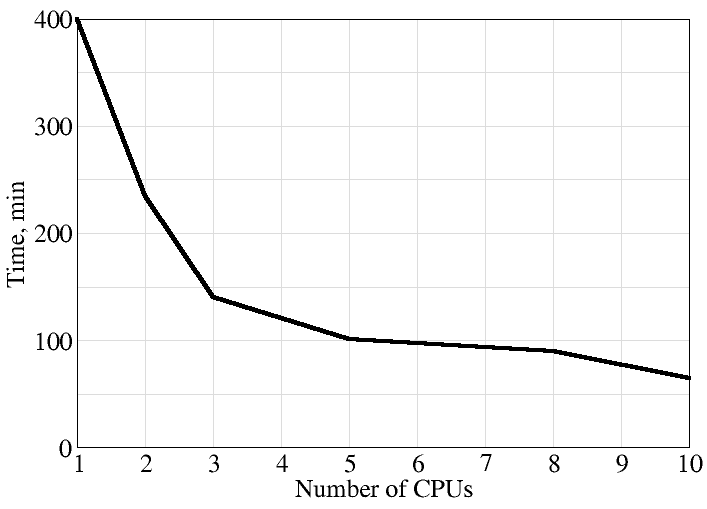}
        \caption{Calculation time $t_{n_p}$ in min}
        \label{fig:time}
    \end{subfigure}
    \hfill
    \begin{subfigure}[b]{0.32\textwidth}
        \centering
        \includegraphics[width=\textwidth]{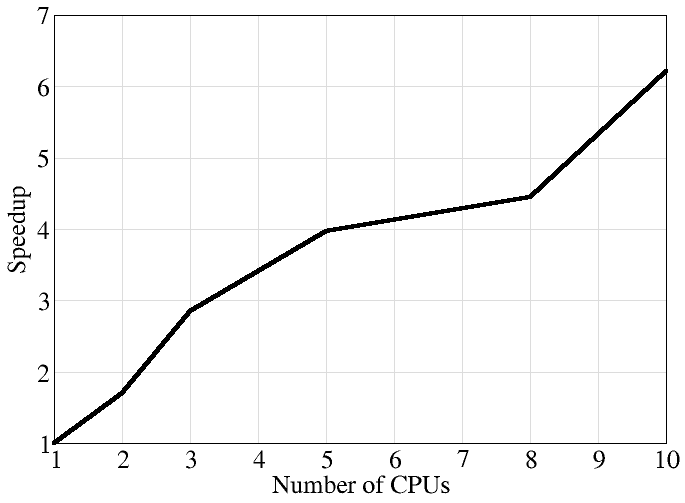}
        \caption{Speedup: $t_{1}/t_{n_p}$}
        \label{fig:speedup}
    \end{subfigure}
    \hfill
    \begin{subfigure}[b]{0.32\textwidth}
        \centering
        \includegraphics[width=\textwidth]{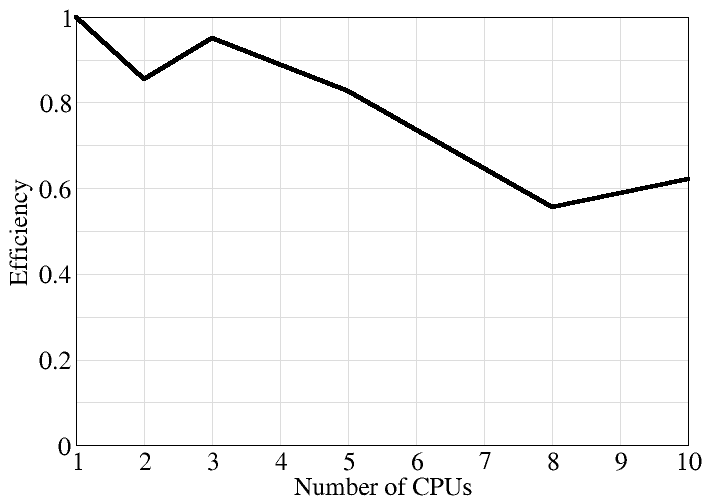}
        \caption{Efficiency:  $t_{1}/\left(n_p\times t_{n_p}\right)$}
        \label{fig:efficiency}
    \end{subfigure}
    \vspace{-5mm}
    \caption{Results of the parallel algorithm: time, speedup and efficiency of calculations in dependence of number of processors -- $n_p$}
    \label{fig:parallel_results}
    \vspace{-8mm}
\end{figure}

Fig.~\ref{fig:parallel_results} illustrates the performance of the parallel algorithm depending on the number of CPUs -- $n_p$. The~calculations have been carried out on HybriLIT heterogenius cluster  (CPU -- Intel Xeon E5-2695). Fig.~\ref{fig:time} shows the calculation time $t_{n_p}$, Fig.~\ref{fig:speedup} -- the speedup of calculations $t_{1}/t_{n_p}$, and Fig.~\ref{fig:efficiency} -- the efficiency of~parallelization $t_{1}/\left(n_p\times t_{n_p}\right)$.

\section{Summary and Conclusion}
\label{summary}
In order to maximize the working area we have suggested changing the plate geometry. A new shape has been derived from the regular one by removal of rectangular parts from its corners. The simulations have been carried out in order to find the optimal plate geometry. For that the algorithm for solving the one-parameter optimization problem has been developed and its parallel version has been implemented.

The optimal geometry of the plate has been found ($l^*=0.6~\mathrm{cm}$) allowing to increase the working area up~to~$23.5\%$ in comparison with the one of the regular plate. The value of the electric current corresponding to the maximum working area has been found and equals to $I=385~\mathrm{mA}$.

\begin{acknowledgement}
The study is partially supported by RFBR according to the projects No.~14-01-00628 and No.~14-01-31227. Authors thank Edik Ayryan (JINR) for permanent interest in this research and worthwhile discussions, and S.~Lebedev (JINR \& Giessen University) for helpful remarks.
\end{acknowledgement}

%
%
%

\end{document}